\documentclass[12pt]{article}
\usepackage[utf8]{inputenc}
\usepackage{amsmath,dsfont,amssymb,bm,amsthm}
\usepackage{epstopdf,geometry,natbib,color}
\usepackage{graphicx,float,subfigure}
\usepackage{float}
\usepackage{hyperref}
\usepackage{soul}
\usepackage{multirow}
\providecommand{\keywords}[1]{\textbf{\textit{Keywords---}} #1}

\def\R{\mathbb R}
\newcommand{\ds}{\displaystyle}

\newcommand{\Ll}{\mathcal{L}}

\definecolor{aquamarine}{rgb}{0.13, 0.68, 0.8}

\begin{document}
	
\title{Mechanistic–statistical inference of mosquito dynamics from mark–release–recapture data}

\author{
Nga Nguyen$^{\footnotesize\hbox{a}}$, 
Olivier Bonnefon$^{\footnotesize\hbox{b}}$, 
Ren\'e Gato$^{\footnotesize\hbox{c}}$, 
Luis Almeida$^{\footnotesize\hbox{d}}$, 
Lionel Roques$^{\footnotesize\hbox{b}}$\\[1ex]
\footnotesize{$^{\hbox{a}}$  LAGA, Institut Galilée, University Sorbonne Paris Nord, 93430 Villetaneuse, France }\\
\footnotesize{$^{\hbox{b}}$  INRAE, BioSP, 84914, Avignon, France}\\
\footnotesize{$^{\hbox{c}}$ Institute of Tropical Medicine Pedro Kourí (IPK), Havana, Cuba}\\
\footnotesize{$^{\hbox{d}}$  Sorbonne Universit\'e, CNRS, Universit\'e de Paris, LPSM, 75005 Paris, France}\\
}
\date{}
\maketitle
	
\begin{abstract}
Biological control strategies against mosquito-borne diseases require reliable field estimates of dispersal and survival of released males. We present a mechanistic–statistical framework for mark–release–recapture (MRR) data that links (i) an individual-based 2D It\^o diffusion model with stochastic death and capture, and (ii) its reaction–diffusion approximation yielding expected densities and trap-specific captures. Inference is achieved by solving the reaction–diffusion system within a Poisson observation model for daily trap counts, with uncertainty assessed via parametric bootstrap. We first validate parameter identifiability using simulated data that closely mimic the experimental data. We then analyze an urban MRR campaign in El Cano (Havana, Cuba) comprising four weekly releases of differentially marked sterile \textit{Aedes aegypti} males and a network of indoor BG-Sentinel traps. A homogeneous mobility model is favored by AIC over an urban–nonresidential heterogeneous alternative. Estimates indicate a post-release life expectancy of about five days (eight days total adult age) and a typical displacement of 180~m after five days, with trap efficiency jointly identified alongside movement and survival. Unlike empirical analyses that summarize trap returns (e.g., exponential fits of daily survival or mean recapture distance), the mechanistic formulation couples movement, mortality, and capture explicitly and reduces biases from trap layout—enabling joint estimation and biologically interpretable parameters. The approach delivers a calibrated, computationally efficient procedure for extracting biologically interpretable parameters from sparse MRR data and offers a principled alternative to empirical summaries for the design and evaluation of sterile insect technique (SIT)-style interventions in urban settings.
\end{abstract}


\section{Introduction}

Mosquitoes are the primary arthropod vectors of human diseases worldwide, transmitting numerous illnesses including malaria, lymphatic filariasis, and arboviruses such as dengue, chikungunya and Zika virus. For several of these diseases, available treatments are only symptomatic, and for most, no widely deployable specific vaccines or drugs currently exist. Under these circumstances, controlling mosquito populations remains the primary means of prevention. Given the limitations of conventional insecticide-based approaches, there is an increasing need for innovative strategies that are both sustainable and environmentally friendly \citep{ACH19,BEC20}. Among these alternatives, biological control methods, such as the sterile insect technique (SIT), the release of insects carrying a dominant lethal (RIDL), or Wolbachia-based strategies, involve the release of large numbers of mosquitoes that are either sterile or less capable of transmitting disease.

\textit{Aedes aegypti} is the primary vector of dengue worldwide. This mosquito lives in close association with humans and shows a strong preference for feeding on human blood. Its larvae can develop in a wide range of water reservoirs associated with household activities \citep{HAM07}. These characteristics make it an ideal vector for the spread of dengue virus, particularly in large cities with dense populations and numerous artificial water containers. In this context, the SIT has been progressively evaluated—from laboratory studies to large-cage experiments—by one of the present authors and collaborators, to systematically assess potential impacts on mosquito performance and survival under increasingly natural conditions \citep{GAT14,GAT21}. However, a critical gap remains: reliable estimation of key ecological parameters of released sterile males (such as dispersal and survival) in real-life urban settings, which are essential to the design and scheduling of release strategies.

To address this gap, we designed a single mark–release–recapture (MRR) campaign, implemented by one of the present authors and collaborators, explicitly to estimate ecological parameters of sterile male \textit{A. aegypti} under field conditions. The campaign consisted of four weekly releases of differentially marked sterile males at a fixed urban site in Cuba and a dense indoor trap network for daily recaptures. 

MRR experiments, combined with spatio-temporal dispersal models, have long been used to infer demographic parameters including survival rates, longevity, and emigration \citep{LebBur92,SIL07,MRR08,GroGim10}. Yet, mosquito MRRs pose distinctive challenges: short lifespans and low recapture rates yield sparse, noisy data \citep{GAR16}, and classical population-scale models can be inadequate. This motivates the use of microscopic individual-based models (IBMs), which represent each mosquito explicitly, but IBMs are computationally intensive and their stochasticity complicates parameter inference \citep{HarCal11}. 

Our methodological contribution is to integrate two complementary modeling frameworks. At the microscopic level, we model mosquito movement as a two-dimensional It\^o diffusion with spatially varying mobility, and we treat death and capture as stochastic events. At the macroscopic level, we derive a reaction–diffusion formulation that is the deterministic limit of the IBM and yields the expected population density and expected captures. This theoretically grounded link allows us to estimate IBM parameters without direct large-scale simulation: inference proceeds by solving the reaction–diffusion system and embedding it in a mechanistic–statistical observation model. Such mechanistic–statistical approaches have proved effective for reaction–diffusion systems under sparse, noisy observations—either when models are constructed from biological principles \citep{RoqSouRou11,RoqBon16} or when they summarize underlying stochastic processes \citep{RoqWal16,RoqSou22}—and naturally fit within a state-space framework \citep{PatTho08,DurKoo12}. A central focus here is the explicit modeling of trapping and the joint identifiability of movement, survival, and trap efficiency.

We first validate the inference procedure on simulated datasets to assess parameter identifiability. We then apply it to the Cuban field campaign reported here, thereby obtaining calibrated estimates of survival and mobility in a real urban context. Our objectives are threefold: (1) to deliver a well-calibrated mechanistic description of sterile male dynamics under field conditions; (2) to quantify how spatial heterogeneity (urban versus non-residential environments) influences mobility; and (3) to propose a generic, transferable estimation procedure for parameters of stochastic (IBM-based) and deterministic diffusion models from MRR data.

We remark that obtaining good estimates on how mosquito spatial diffusion depends on the heterogeneity of the terrain is important to be able to calibrate mathematical models that take into account spatially heterogeneous environments in order to control mosquito populations in an efficient and robust way \citep[see, for instance,][]{AgboAlmeidaCoron2}.

\section{Material and Methods}
	\label{sec:methods}
\subsection{Mark-release-recapture data}
	\label{subsec:data}
 
%\Rd Le paragraphe suivant parle surtout de SIT; ici ce n'est pas le point important et tel que c'est écrit, on ne le comprend pas. Il faudrait peut etre plutot écrire que cette expérience de MRR a été réalisée dans un cadre plus général d'évaluation de la SIT. Il faudrait aussi et surtout etoffer la description de l'expérience de MRR, en incluant les différents détails (dates, effet de la date sur la dynamique du moustique, point exact de lacher, type exact de piège et caractéristiques, si les individus sont retirés chaque jour, comment sont faits les relevés ...)  

The mark-release-recapture (MRR) experiment described in this paper was conducted at a time when wild \textit{A. aegypti}
mosquito populations at the study site exhibited low abundance due to a preceding SIT pilot
study \citep{GAT21}.

\begin{figure}
		\centering
		\includegraphics[width = \textwidth]{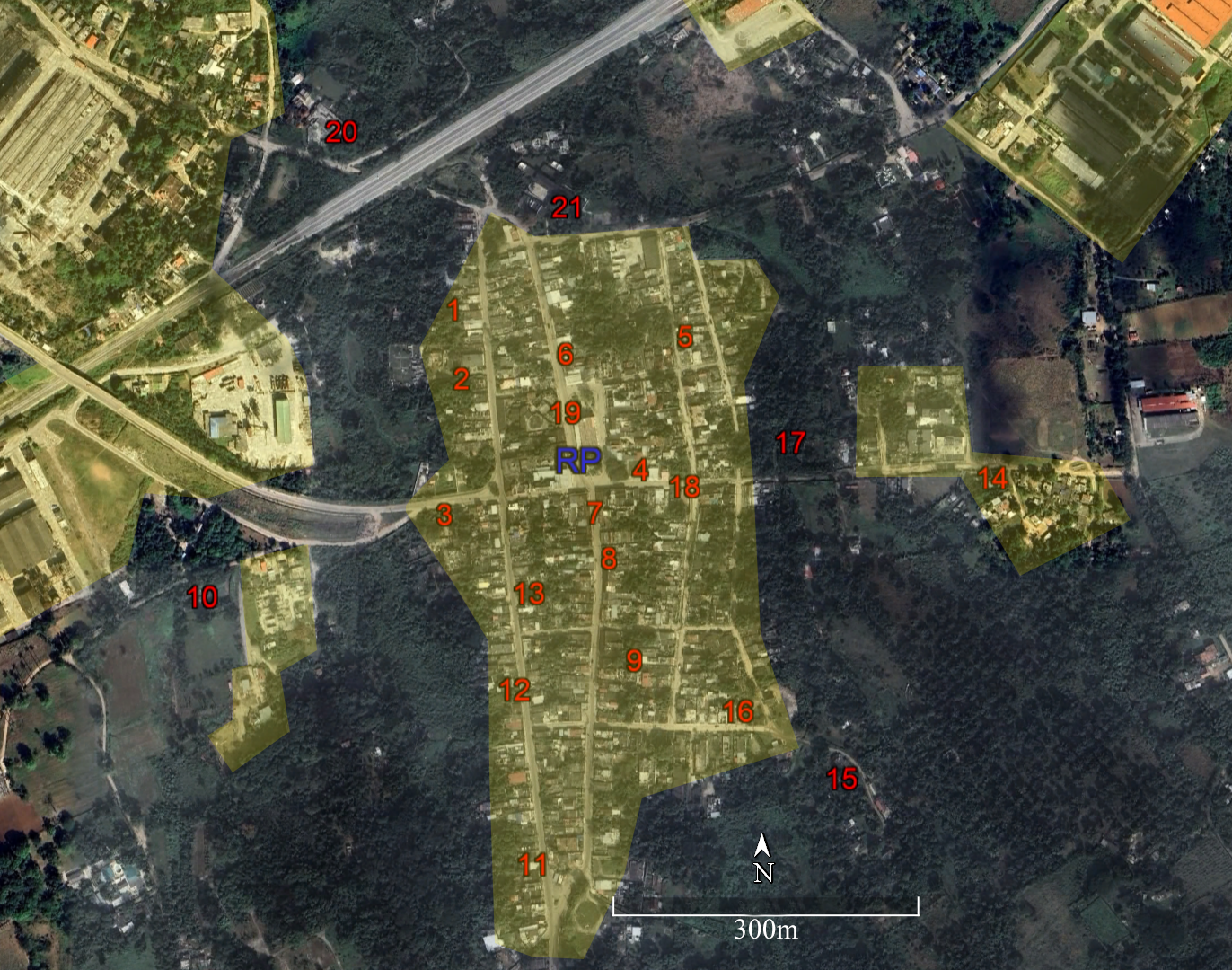}
		\caption{Satellite view of the El Cano study site (Havana, Cuba). The blue marker indicates the release point; red numbers label the 21 BG-Sentinel traps used for daily recaptures. Urban areas are shaded in yellow, where a distinct diffusion coefficient is assumed in the heterogeneous model. Imagery: Google Earth, © Google 2025; data: © Airbus.}
		\label{fig:trap}
	\end{figure}

\noindent\textbf{Study site: }The study was conducted in El Cano, a suburban neighborhood in southwestern
Havana (23°02'00.2"N, 82°27'33.1"W), see Fig.~\ref{fig:trap}. The area spans 50 hectares, with 3,805 residents in 906
houses. The houses are typically small, single-story, and often include backyards with fruit trees.
The experiment took place during the summer, marked by sunshine, rainfall, and thunderstorms,
with light to moderate winds (15–25 km/h). El Cano is relatively isolated from the rest of the
metropolitan area by rural regions, forests, and infrastructure, which limits mosquito migration.

\noindent\textbf{Study Design: }The study was designed following the guidelines described by \cite{Bouyer24}. Four releases of 10,000 marked mosquitoes each were carried out over consecutive weeks
at a single location, referred to as the release point, at the center of the study site. Each week,
the mosquitoes were marked with a different color using a fluorescent powder: yellow, red, blue, and pink, respectively. An
adult mosquito trap network was deployed to recapture the released mosquitoes and monitor
the wild population. The network consisted of twenty-one BG-Sentinel traps baited with BG-lures
(Biogents, Germany), distributed along concentric rings at distances of approximately 50, 100, 150, 200, 250, 300 and
400 meters. The traps were installed indoors, at ground level, in the quietest areas of
inhabited buildings, and were checked daily for six weeks. The collected biological material was
transported to the laboratory (at IPK) in plastic containers, where insects were killed by freezing at -20°C.
Mosquitoes were then identified morphologically by species and sex under a stereomicroscope.
Males were further classified as either wild, unmarked individuals or marked with different dust
colors, using ultraviolet light for identification.

\noindent\textbf{Mosquitoes: }\textit{A. aegypti} were colonized from eggs collected in ovitraps at the study site in
2018. Adults were reared in cages at a density of one mosquito/cm², with a 1:2 male-to-female
ratio, and fed a 10\% honey solution. The colony was maintained at 28 ± 2°C and 80 ± 10\%
relative humidity. Porcine blood at 38°C was provided in collagen casings (Fibran, Girona, Spain)
once a week for female feeding. Eggs were collected 2–3 days post female feeding and allowed to
mature in a wet environment for 3 days. Larvae were reared in 100 cm × 60 cm × 3 cm trays,
each containing 4 L of deionized water, at a density of 2 larvae/mL. Larval density was
determined using egg quantity-weight regression curves as described in \citep{zheng2015standard}.
Simultaneous hatching was induced by immersing eggs in 36°C de-oxygenated water under
vacuum. First-instar larvae were then transferred to trays. The IAEA standard diet (50\% tuna
meal, 36\% bovine liver powder, and 14\% brewer’s yeast) was provided daily at the rates of 0.2,
0.4, 0.8, and 0.6 mg/larva for larval instars I, II, III, and IV, respectively \citep{Puggioli2013}. After the onset of
pupation, immatures were collected and sorted by sex using a Fay-Morlan apparatus (John W.
Hock Co, Gainesville, FL, USA). Male pupae were dosed by volume into 10 mL plastic tubes, each
containing approximately 500 individuals. Batches of 5,000 male pupae were kept in 1 L flat
tissue-culture flasks (Thermo Fisher, Waltham, MA, USA) filled with 250 mL of dechlorinated
water until the optimal age for irradiation was reached. The flasks were placed horizontally to
maximize the water surface area.

\noindent\textbf{Sterilization and Packing: }Male mosquito pupae were irradiated using a 60Co Isogamma
LLCo irradiator (Izotop, Budapest, Hungary) as close to emergence as possible to minimize
somatic damage, i.e., aged at least 30 hours. An irradiation dose of 80 Gy was applied with a
dose rate of 8 kGy/h. After irradiation, the pupae were put in the culture flasks for
transport and emergence. Cardboard boxes (15 cm × 15 cm × 60 cm) were placed horizontally
and used as adult containers. Square holes (10 cm × 10 cm) were cut in the two smaller sides of
the box and covered with fine mesh secured with a rubber band. Additionally, a 3 cm diameter
hole was made on one of the 15 cm × 60 cm sides of the box, so that the neck of a culture
flask with 5,000 irradiated male pupae could be introduced. Emerged adult mosquitoes moved naturally through the neck of the flask, into the box which served as a resting area. Once all adults had finished emerging, the
flasks were removed and the holes were covered with a 50 mL plastic tube coated with honey-soaked filter paper.

\noindent\textbf{Marking: }The boxes containing two-day-old adult sterile males were placed in a fridge at 4°C
for 15 minutes to immobilize the mosquitoes. The immobilized mosquitoes were transferred in
batches of approximately 2,500 specimens to 1-liter plastic containers containing 10.4 mg of
fluorescent powder (DayGlo® Color Corp., USA) following previously described procedures
\citep{Bouyer24}. The containers were gently
rotated for ten seconds to ensure all mosquitoes came into contact with the pigment. Marked
mosquitoes were then transferred to 30 x 30 x 30 cm metallic cages (BioQuip, USA). Mosquitoes
were provided with water and honey solution for 24 hours prior to release.

\noindent\textbf{Releases: }Sterile mosquitoes were released shortly after sunrise (around 7:00 a.m.), when
temperature and humidity were generally favorable (temperature ranging from 22.1 to 26.4°C
and humidity from 72 to 93\%). Mosquitoes were released as 3-day-old adults by opening the lid
of the boxes.

The numbers of mosquitoes captured each day in each trap were stored as tables (see Appendix~1). Four datasets correspond to the four different releases that were conducted. In this work, in the absence of environmental data distinguishing one release experiment from another, we consider these four datasets as four experimental replicates. Since the releases were spaced only one week apart, the experiments overlap in time. However, the use of different marking colors ensures that the mosquitoes captured can be accurately identified with the release (and therefore the experiment) they belong to.

The four datasets are denoted as $\hbox{Obs}^k := \{\hat{y}_i^{j,k}, \ i=1,\ldots,21, \ j=0,\ldots,19\}$, where $\hat{y}_i^{j,k}$ represents the number of mosquitoes captured in trap $i$ on day $j$. The index $k$ corresponds to each of the four replicate MRR experiments.

	%%%%%%%%%%%%%%%%%%%%%%%%%%%%%%%%%%%%%%%%%%%%%%%%%%%%%%%%%%%%%%%%%%%%%%%%
\subsection{Mechanistic models and simulated data \label{subsec:models}}
\paragraph{Microscopic model.}
We describe the mosquito dynamics in the mark-release-recapture analysis by an individual-based model as follows. 

\

\noindent \textit{Movement.} At the start of the release experiment, there are $N_0=10^4$ mosquitoes, all of which are released at the same point $x_0 \in \mathbb{R}^2$. We assume that their positions at time $t$ are governed by independent 2-dimensional It\^o diffusion processes:
\begin{equation}
    dX_t = \sigma(X_t) dB_t, \quad X_0 = x_0,
    \label{eqn:process}
\end{equation}
where $dB_t$ denotes an increment in time of a standard 2-dimensional Brownian motion \citep{Gar09}, and $\sigma$ is a Lipschitz-continuous function on $\mathbb{R}^2$ that describes the local mosquito mobility, which may vary depending on local conditions.

	\
	
	\noindent \textit{Life expectancy and death times.} In the absence of trapping, the mosquito's life expectancy is given by $1/\nu>0$. Their death times follow an exponential distribution with parameter $\nu$.
	
	\

	\noindent \textit{Trapping.} The traps, indexed by $i=1,\ldots,21$, are located at positions $q_i \in \mathbb{R}^2$. For a mosquito at position $X_t$, the probability of being trapped in trap $i$ follows an exponential distribution with the parameter $f_i(X_t) = \gamma \, \exp\left(-\|X_t - q_i\|^2 / R^2\right)$. This implies that, for an immobile individual at position $x$, the average time before being captured is $\min\limits_{i=1,\ldots,21}\left(1 / f_i(x)\right)$. The constant $R > 0$ (assumed to be $R = 10$~m) represents the characteristic distance over which the trap's effectiveness diminishes with increasing distance from $q_i$. At a distance of $R$, the trap's effectiveness is $\gamma\exp(-1)$, which corresponds to $37\%$ of its maximum value.
 The parameter $\gamma$, which is to be estimated, represents the baseline trapping rate for an individual located exactly at the trap position (in such a case, the individual would be trapped after an average duration of $1/\gamma$).

\paragraph{Macroscopic Model.}
First, ignoring the dead or alive status of the mosquitoes, the probability density $v(t,x)$ of the random variable $X_t$ is given by the standard Fokker-Planck equation \citep{Gar09}:
\begin{equation}
    \label{eqn:pde_v}
    \left\{
    \begin{array}{rl}
        \ds \frac{\partial v}{\partial t} & = \ds \Delta\left(\frac{\sigma(x)^2}{2} v\right), \quad t > 0, \quad x \in \mathbb{R}^2, \\
        \ds v(0,x) & = \ds \delta_{x=x_0},
    \end{array}\right.
\end{equation}
where $\Delta$ denotes the standard Laplace operator, and $\delta_{x=x_0}$ represents a Dirac mass at $x = x_0$.

Next, for any given mosquito, we take into account the \emph{killing mechanism}: the process ``dies" at a random time $\tau$ with a hazard rate $F(X_t) := \nu + \sum\limits_{i=1}^{21} f_i(X_t)$: if the process is alive at time $t$, the infinitesimal probability of dying in the next instant $dt$ is $F(X_t)\, dt + o(dt)$.

Let $p(t,x)$ denote the probability density function of the mosquito being alive at time $t$ and located at $x$.
The killing mechanism removes probability mass at a rate $F(x) p(t,x)$. Thus, the corresponding Fokker-Planck equation is given by \citep[see, e.g.,][]{RoqSou22}:
\begin{equation}
    \label{eqn:pde_p}
    \left\{
    \begin{array}{rl}
        \ds \frac{\partial p}{\partial t} & = \ds \Delta\left(\frac{\sigma(x)^2}{2} p\right) - p\, \nu  - p \sum\limits_{i=1}^{21} f_i(x), \quad t > 0, \quad x \in \mathbb{R}^2, \\
        \ds p(0,x) & = \ds \delta_{x=x_0}.
    \end{array}\right.
\end{equation}

We then define $\pi_i(t)$ as the probability that the mosquito has been trapped in trap $i$ before time $t$. This probability satisfies the following equation:
\begin{equation}
    \label{eqn:pi}
    \pi_i'(t) = \int_{\mathbb{R}^2} f_i(x) \, p(t,x) \, dx.
\end{equation}

Finally, since the mosquitoes are assumed not to interact, the expected mosquito population density $h(t,x)$ is the solution to the following equation:
\begin{equation} \label{eq:EDP_pop_dens}
    \left\{
    \begin{array}{rl}
        \ds \frac{\partial h}{\partial t} & = \ds \Delta\left(\frac{\sigma(x)^2}{2} h\right) - h\, \nu  - h \sum\limits_{i=1}^{21} f_i(x), \quad t > 0, \quad x \in \mathbb{R}^2, \\
        \ds h(0,x) & = \ds N_0 \, \delta_{x=x_0}.
    \end{array}\right.
\end{equation}

\begin{figure}
\begin{center}
\includegraphics[scale=0.4]{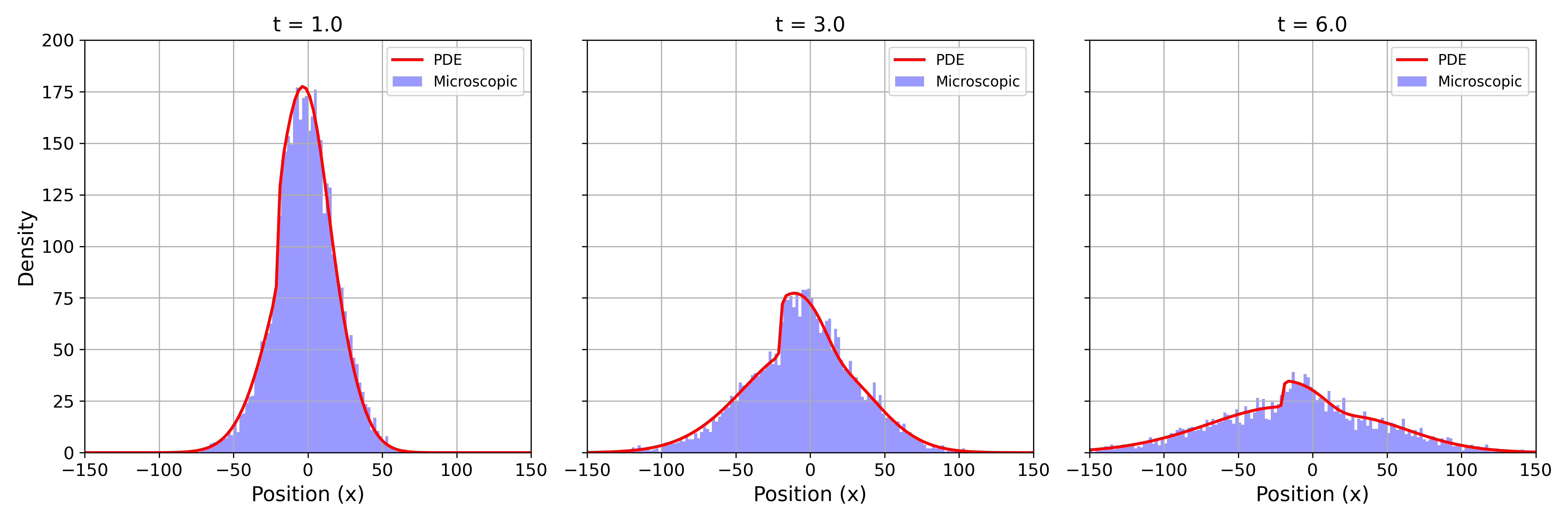}
\end{center}
\caption{Solution $h(t,x)$ of the PDE~\eqref{eq:EDP_pop_dens} vs population density obtained from the microscopic model, in a simplified one-dimensional case with a single trap located at $x=q_1=10$. The mobility coefficient is $\sigma(x)=  \sigma_1 + (\sigma2 - \sigma1)  (1 + \tanh(x + 20))/2$. The parameter values are $x_0=0$, $N_0=10^4$,  $1/\nu=$10~days, $1/\gamma=$2~days, $\sigma_1 = 25~\text{m}/\sqrt{\text{day}}$, $\sigma_2 = 20~\text{m}/\sqrt{\text{day}}$. } 
\label{fig:1D}
\end{figure}

For illustration, in Fig.~\ref{fig:1D} we compare the dynamics of the population densities given by the microscopic and macroscopic models in a simplified one-dimensional case with a single trap. The results are consistent. Moreover, the expected number of trapped individuals obtained from Eqs.~\eqref{eqn:pde_p}-\eqref{eqn:pi} closely matches the values from the microscopic model (see Appendix~2).

\paragraph{Unknown parameters.} The microscopic and macroscopic models are characterized by the same parameters and coefficients: $x_0$ (release point), $N_0$ (number of released mosquitoes), $\nu$ (inverse of post-release life expectancy), $\gamma$ (trapping rate), $R$ (characteristic distance associated with trap effectiveness), and $\sigma(x)$ (mobility coefficient). The first two parameters, $x_0$ and $N_0$, are known, while the others are not, and their estimation will be the main challenge of this work. Before proceeding further, we note that if the quantity $p(t,x)$ in the macroscopic model is approximated by a spatially constant function $P(t)$ in a given trap $\omega_i$, then $\pi_i'(t) = P(t) \, \gamma \, \pi \, R^2$ depends only on the product $\gamma \, R^2$. Thus, the two parameters $\gamma$ and $R$, which describe the capture process, are likely not identifiable, and one of them must be fixed. Here, we fix $R = 10$~m.

Regarding the unknown coefficient $\sigma(x)$, we consider two scenarios: in the first case (Homogeneous case), we assume that $\sigma(x) = \sigma>0$ is constant across all regions. In the second case (Heterogeneous case), we assume that mosquito mobility differs between urban areas and forest areas, denoted respectively by $\Omega_1$ and $\Omega_2$ (see Fig.~\ref{fig:trap}; for mathematical convenience, we assume that $\Omega_1 \cup \Omega_2 = \mathbb{R}^2$). In this case, we assume that $\sigma(x)$ is a spatial regularization of the piecewise constant function
$$
s(x) := 
\begin{cases}
\sigma_1>0 & \text{if } x \in \Omega_1, \\
\sigma_2>0 & \text{if } x \in \Omega_2.
\end{cases}
$$
Specifically,
\begin{equation} \label{eq:sigma_x}
\sigma(x) = \int_{\mathbb{R}^2} J(x-y) \, s(y)\, dy,    
\end{equation}
where $J$ is a Gaussian kernel with a fixed variance (standard deviation =$10$ meters). This regularization procedure is a straightforward method to ensure the well-posedness of the solutions to both the microscopic and macroscopic models.

Finally, we denote by $\Theta$ the vector corresponding to the unknown parameters. In the homogeneous case, we have  $\Theta:= (\sigma, \nu, \gamma)$, and in the heterogeneous case, $\Theta := (\sigma_1, \sigma_2, \nu, \gamma)$.

\paragraph{Simulated data.} Using the microscopic model described above, we generated datasets that mimic real MRR (mark-release-recapture) data. These datasets serve as test sets to validate our estimation procedure. 
Specifically, we fixed the unknown parameters $\Theta_{\text{true}} := (\sigma_{\text{true}}, \nu_{\text{true}}, \gamma_{\text{true}})$ ($\Theta_{\text{true}} := (\sigma_{1,{\text{true}}},\sigma_{2,{\text{true}}}, \nu_{\text{true}}, \gamma_{\text{true}})$ in the heterogeneous case) and used the real release position $x_0$, the real trap positions $q_i$ (see Section~\ref{subsec:data}), and the number of released individuals $N_0 = 10^4$ to generate four datasets (in each scenario: Homogeneous and Heterogeneous) describing the number of mosquitoes trapped in each trap each day. Each dataset is of the form
$\tilde{\hbox{Obs}}^k=\{\tilde{y}_i^j, \ i=1,\ldots,21, \ j=0,\ldots,19\}$, where $\tilde{y}_i^j$ denotes the number of mosquitoes trapped in trap $i$ during day $j$.

We performed these simulations using the following parameter values: $1/\nu_{\text{true}} = 10~\text{days}$ and $1/\gamma_{\text{true}} = 1.5~\text{days}$. For the coefficient $\sigma_{\text{true}}(x)$, we considered two cases: i) (Homogeneous case)  $\sigma$ is constant across all regions  and $\sigma_{\text{true}} = 19~\text{m}/\sqrt{\text{day}}$; ii) (Heterogeneous case) mosquito mobility differs between urban areas and forest areas and $\sigma(x)$ is defined by \eqref{eq:sigma_x} ; we set $\sigma_{1,\text{true}} = 50~\text{m}/\sqrt{\text{day}}$ and $\sigma_{2,\text{true}} = 15~\text{m}/\sqrt{\text{day}}$.

\subsection{Parameter estimation with the mechanistic-statistical approach}
\label{subsec:meca-stat}
\paragraph{Observation model.}  A key aspect of mechanistic-statistical modeling is establishing the connection between the available data and the process described through a mechanistic model. This connection is made by deriving a probabilistic model for the observation, conditional on the state of the mechanistic model. Here, the datasets $\hbox{Obs}^k:=\{\hat{y}_i^{j,k}, \ i=1,\ldots,21, \ j=0,\ldots,19\}$ correspond to the number of mosquitoes trapped in trap $i$ during day $j$ (i.e., for $t\in [j,j+1)$). The index $k$ corresponds to each of the four replicate MRR experiments.  Based on the microscopic model described above, each day, any mosquito is trapped in $\omega_i$ with a probability $\pi_i(j+1)-\pi_i(j)$. Thus, $\hat{y}_i^{j,k}$ can be viewed as the sum of $N_0$ independent Bernoulli trials. Since $\pi_i(j+1)-\pi_i(j)\ll 1$ (the probability for a given mosquito to be trapped in trap $i$ during day $j$ is small) and $N_0\gg 1$, the Poisson limit theorem leads to the following observation model for the number of mosquitoes captured in the trap $i$ during day $j$:
	\begin{equation}
		Y_i^j| \Theta \sim \text{Poisson}\left[N_0(\pi_i(j+1)-\pi_i(j))\right],
	\end{equation}
	with $\Theta$ the vector of unknown parameters.  We recall that, although not explicitly stated, $\pi_i(j+1) - \pi_i(j) = \int_{j}^{j+1}\int_{\mathbb{R}^2} f_i(x) \, p(t,x) \, dx \, dt$ depends on $\Theta$, as $p(t,x)$ is the solution of~\eqref{eqn:pde_p} with parameters defined by $\Theta$.

\paragraph{Likelihood function.}    
	Assuming that the observations are independent, conditionally on the diffusion-mortality-capture process, the likelihood associated with  $\Theta$ is:
	\begin{align}
		\label{eqn:likelihood}
		\ds \Ll(\Theta):= P(\hbox{Obs}|\Theta) & = \prod_{k=1}^{4}  \prod_{j=0}^{  19}\prod_{i=1}^{21} P(Y_i^j=\hat{y}_i^{j,k})\nonumber \\
		& = \prod_{k=1}^{4} \prod_{j=0}^{19}\prod_{i=1}^{21} \exp[-N_0(\pi_i(j+1)-\pi_i(j))]\frac{\left[N_0(\pi_i(j+1)-\pi_i(j))\right]^{\hat{y}_i^{j,k}}}{\hat{y}_i^{j,k}!}.
	\end{align}
In practice, we computed the log-likelihood:
	\begin{align}
		\label{eqn:log-likelihood}
		\ds \ln \Ll(\Theta) & = \sum_{k=1}^{4}\sum_{j=0}^{19}\sum_{i=1}^{21} [-N_0(\pi_i(j+1)-\pi_i(j))] + \hat{y}_i^{j,k} \ln \left[N_0(\pi_i(j+1)-\pi_i(j))\right] - \ln(\hat{y}_i^{j,k}!) \nonumber \\ & = \hat{C} - 4\, N_0 \sum_{i=1}^{21} \pi_i (20) + \sum_{k=1}^{4}\sum_{j=0}^{19}\sum_{i=1}^{21} \hat{y}_i^{j,k} \ln \left[N_0(\pi_i(j+1)-\pi_i(j))\right],
	\end{align}
with $\hat{C}$ a constant which does not depend on $\Theta$.

\paragraph{Computation of the maximum likelihood estimator and confidence intervals.}  The maximum likelihood estimator (MLE) is defined as the parameter $\Theta^*$ that maximizes the log-likelihood $\ln \Ll (\Theta)$ among the admissible values of $\Theta$, specifically $\Theta \in (0,\infty)^3$ in the homogeneous case or $\Theta \in (0,\infty)^4$ in the heterogeneous case. The MLE is computed using the BFGS minimization algorithm, applied to $-\ln \Ll(\Theta)$. 

The computation of the standard deviations for the estimated parameters relies on a parametric bootstrap built on the microscopic (IBM) model. After obtaining the MLE $\Theta^*$ from the simulated or experimental data, we proceed as follows: we generate $B=100$ independent synthetic datasets $\{\mathrm{Obs}^{(b)}\}_{b=1}^B$ using the IBM with parameter $\Theta^*$ and the same experimental design (release size, trap locations, observation window). For each bootstrap dataset, we re-estimate the parameters to obtain $\Theta^{(b)}$. The standard deviation for each component $\Theta_i$ is $\mathrm{Std}(\Theta_i)=\sqrt{\bigl[\widehat{\Sigma}\bigr]_{ii}}$, with $\widehat{\Sigma}$ the empirical covariance matrix.

\paragraph{Parameter bounds.}  
For the estimation using the constrained BFGS algorithm, we need some a priori bounds on the parameter values.
Regarding the mobility parameter, at each position $x$, the expected mean squared displacement of $X_t$ during a small time interval $\tau$ is 
$$
\mathbb{E}(\|X_{t+\tau} - X_t\|^2 \,|\, X_t = x) \approx 2 \, \sigma(x)^2 \, \tau.
$$
We take $\tau = 1/1440~\text{day}$ (1 minute). To determine a lower bound, we assume that the expected mean squared displacement during $\tau$ is $0.01~\text{m}^2$, and for an upper bound, we assume it is $100~\text{m}^2$. Regarding the parameter $\nu$, we use a minimal life expectancy of 1 day and a maximum life expectancy of 50 days. Regarding the parameter $\gamma$, we assume that the mean duration before capture at $x = q_i$ (the center of a trap) is between 1 minute and 10 days. Finally this leads to 
$$
\sigma \in (2.7, 268),\ \nu \in (0.02,1), \ \gamma \in( 0.1,1440).
$$

\subsection{Numerical and software aspects}

\paragraph{1D PDE toy model.} 
The reaction--diffusion model is discretized in space using second-order finite differences on a uniform grid with homogeneous Neumann (no-flux) boundary conditions. This yields a semi-discrete ODE system (method of lines). Time integration is performed with the \texttt{odeint} solver from \textsc{SciPy}, which relies on \textsc{LSODA} to adaptively switch between Adams and \textsc{BDF} schemes. A Jupyter notebook is available \href{https://doi.org/10.5281/zenodo.17076370}{here}.

\paragraph{2D PDE model.} 
We solve the reaction--diffusion model on the square domain $\Omega=[-1000,1000]^2$ (meters) with homogeneous Neumann (no-flux) boundary conditions. Spatial discretization uses finite elements, with implicit time stepping for the diffusion term and Crank--Nicolson for the source terms (mortality and trapping). Because accuracy depends on the discretization, the mesh (and, when necessary, the time step) is refined until the computed number of captured mosquitoes stabilizes. A picture of the triangular mesh is available in Appendix~3. For optimization, we use a BFGS algorithm that requires both the objective $-\ln\mathcal{L}(\Theta)$ and its parameter derivatives; these derivatives are obtained with a linear tangent model, solved once for each component of $\Theta$ using the same discretization and boundary conditions.  

The 2D MRR simulator is implemented in Python and builds upon the MSE software \href{https://mse.biosp.org/}, which is dedicated to the simulation and calibration of PDE models. The MRR simulator is available in a dedicated Git repository, \href{https://forge.inrae.fr/msegrp/mrr-simulator}{here} [https://forge.inrae.fr/msegrp/mrr-simulator]. Since MSE is distributed through GUIX \citep{courtes:hal-04586520}, this work can be reproduced in a controlled environment on Linux operating systems.

\paragraph{Microscopic model.} 
A detailed algorithmic description of the simulation procedure is provided in Appendix~4, and is also available in a supplementary Jupyter notebook \href{https://doi.org/10.5281/zenodo.17076370}{here}.

\section{Results}
\label{sec:results}

\subsection{Validation of the estimation procedure on the simulated dataset}
\label{subsec:simudata}

\begin{table}
 	\caption{Validation of the estimation procedure. Maximum-likelihood estimates and parametric-bootstrap standard deviations (Std) for the homogeneous ($\sigma,\ \nu,\ \gamma$) and heterogeneous ($\sigma_{1},\ \sigma_{2},\ \nu,\ \gamma$) models fitted to simulated MRR data.}
 	\centering
 	\begin{tabular}{l c c c c }
 		\hline
 		\multicolumn{5}{c}{\bf Homogeneous case} \\
 		\hline
 		Parameter $\Theta$&  \multicolumn{2}{c}{$\sigma$} & $\nu$ & $\gamma$ \\
 		\hline
 		$\Theta_{\text{true}}$ &  \multicolumn{2}{c}{19.00} & 0.100 & 0.667  \\
 		%MLE & \multicolumn{2}{c}{17.4663} & 0.1391 & 0.8194 \\
 		MLE $\Theta^*$ & \multicolumn{2}{c}{18.83} & 0.100 & 0.654 \\
$\mathrm{Std}(\Theta^*)$ & \multicolumn{2}{c}{$1.48e^{-1}$} & $2.0e^{-3}$ & $1.76e^{-2}$  \\
        \hline
 		\hline
 		\multicolumn{5}{c}{\bf Heterogeneous case} \\
        \hline
 		Parameter $\Theta$& {$\sigma_1$} & $\sigma_2$& $\nu$ & $\gamma$ \\
 		\hline
 		$\Theta_{\text{true}}$ & {50.00} &15.00& 0.100 & 0.667 \\
 		%MLE & \multicolumn{2}{c}{17.4663} & 0.1391 & 0.8194 \\
 		MLE $\Theta^*$ & {50.20}& 14.67 & 0.100 & $0.674$ \\
        %$\mathrm{Std}(\Theta)$& 27.05& 20.88 & $2.329e^{-3}$ & $1.275e^{-2}$ & \\
        $\mathrm{Std}(\Theta^*)$ & $4.44e^{-1}$& $1.00$ & $2.89e^{-3}$ & $1.50e^{-2}$ \\
    \end{tabular}
 	\label{tab:result_simu}
 \end{table}

\paragraph{Homogeneous case.} We estimated the parameters by maximum likelihood (MLE) using the simulated dataset with true values $\Theta_{\text{true}}=(\sigma_{\text{true}},\nu_{\text{true}},\gamma_{\text{true}})=(19,\,0.10,\,2/3)$.
The results in Table~\ref{tab:result_simu} show good agreement between $\Theta_{\text{true}}$ and the MLE $\Theta^*$.
To quantify this, we computed the z-score $z(\theta)=|\theta^*-\theta_{\text{true}}|/\mathrm{Std}(\theta^*)$:
$z(\sigma)=1.15$, $z(\nu)=0.00$, $z(\gamma)=0.72$.
Thus, $\nu$ and $\gamma$ are within one Std and $\sigma$ is only slightly above (relative error $\approx 0.9\%$), confirming accurate estimation.

\paragraph{Heterogeneous case.} The same procedure was applied with a heterogeneous mobility coefficient $\sigma(x)$. In this case, the simulated dataset corresponds to the true parameter $\Theta_{\text{true}} = (\sigma_{1,\text{true}}, \sigma_{2,\text{true}}, \nu_{\text{true}}, \gamma_{\text{true}}) = (50, 15, 1/10, 2/3)$. The results are presented in Table~\ref{tab:result_simu}.
Using the same definition, we obtained $z(\sigma_1)=0.45$, $z(\sigma_2)=0.33$, $z(\nu)=0.00$, $z(\gamma)=0.47$.
All parameters satisfy $z\le 1$ (within one Std), indicating very good agreement with the true values and supporting strong identifiability.

\subsection{Parameter estimation using experimental data}
We now apply the estimation procedure to the experimental data presented in Section~\ref{subsec:data}. We compare two models: the model with a homogeneous mobility coefficient $\sigma$ and the model with a heterogeneous mobility coefficient $\sigma(x)$. The results are presented in Table~\ref{tab:result_expe}. Since the first model is a particular case of the heterogeneous model (corresponding to $\sigma_1 = \sigma_2 = \sigma$), the heterogeneous model, as expected, leads to a larger log-likelihood associated with the MLE.

To better compare these two models, we use the Akaike Information Criterion (AIC), which introduces a penalty term for the number of parameters in the model \citep{Aka74}:
$
AIC = -2 \ln(\mathcal{L}(\Theta^*)) + 2 N_p,
$
where $\mathcal{L}(\Theta^*)$ is the maximum likelihood of the model, and $N_p$ is the number of parameters in the model ($N_p=3$ in the homogeneous model, $N_p=4$ in the heterogeneous model). 
The AIC balances the goodness of fit (measured by $-2 \ln(\mathcal{L}(\Theta^*))$) and model complexity (penalized by $2 N_p$). A lower AIC value indicates a better model. 

Using the MLE negative log-likelihoods in Table~\ref{tab:result_expe}, we compute:
\[
\mathrm{AIC}_{\text{hom}} = 2\times 293.66 + 2\times 3 = 593.32,\qquad
\mathrm{AIC}_{\text{het}} = 2\times 293.11 + 2\times 4 = 594.22.
\]
Hence the homogeneous model has the lowest AIC. Both models are plausible, but parsimony favors the homogeneous model. With this model, the parameters can be interpreted as follows (see the paragraph on ``Parameter bounds" in Section~\ref{subsec:meca-stat} for a detailed explanation):

    \begin{itemize}
		\item $\sigma = 64.00$: the length of a
		the one-minute straight-line move is about $\lambda = 2.39$ (m). The expected distance to the release point at time $t$ is $
\mathbb{E}(\|X_t\|) = \sigma \sqrt{\tfrac{\pi t}{2}} \approx 80 \sqrt{t}$ (m);
		\item $\nu = 0.21$: the post-release life expectancy of a mosquito is about $5$ days. Including the 3 days prior to release, the total life expectancy is about 8 days. The expected distance from the release point after $5$ days is $\approx 180$ (m);
		\item $\gamma = 0.14$: during each hour a mosquito stays at the position of the trap, there are $\gamma/24 \approx 0.6 \%$ that it is captured. 
	\end{itemize}

\begin{table}
		\caption{Maximum-likelihood estimates and parametric-bootstrap standard deviations (Std) for the homogeneous ($\sigma,\ \nu,\ \gamma$) and heterogeneous ($\sigma_{1},\ \sigma_{2},\ \nu,\ \gamma$) models fitted to the experimental MRR data. The last column reports the negative log-likelihood at the MLE.}
		\centering
		\begin{tabular}{l | c c c c | c}
			\hline
			\multicolumn{6}{c}{\bf Homogeneous case} \\
			\hline
			Parameter $\Theta$ &  \multicolumn{2}{c}{$\sigma$} & $\nu$ & $\gamma$ & $-\ln \mathcal{L}(\Theta^*)$ \\
			\hline
			MLE $\Theta^*$ & \multicolumn{2}{c}{64.00} & $2.104e^{-1}$ & $1.423e^{-1}$  & 293.66\\
            $\mathrm{Std}(\Theta^*)$& \multicolumn{2}{c}{$1.5$ }&$1.71e^{-2}$& $6.8e^{-3}$   & \\
			\hline
			\hline
			\multicolumn{6}{c}{\bf Heterogeneous case} \\
			\hline 
			Parameter $\Theta$ & $\sigma_1$ & $\sigma_2$ & $\nu$ & $\gamma$ & $-\ln \mathcal{L}(\Theta^*)$ \\
			\hline
			%MLE & 63.3800 &  71.5731 &  0.2156 & 0.1434 \\
			MLE $\Theta^*$&64.24& 76.29 &   0.212   & 0.147
            &293.11\\
             $\mathrm{Std}(\Theta^*)$&1.59 &8.74  &$1.28e^{-2}$   &$8.03e^{-3}$ &\\
			\hline 
		\end{tabular}
		\label{tab:result_expe}
	\end{table}

\begin{figure}
	\centering
    \subfigure[]{\includegraphics[width = 0.49\textwidth]{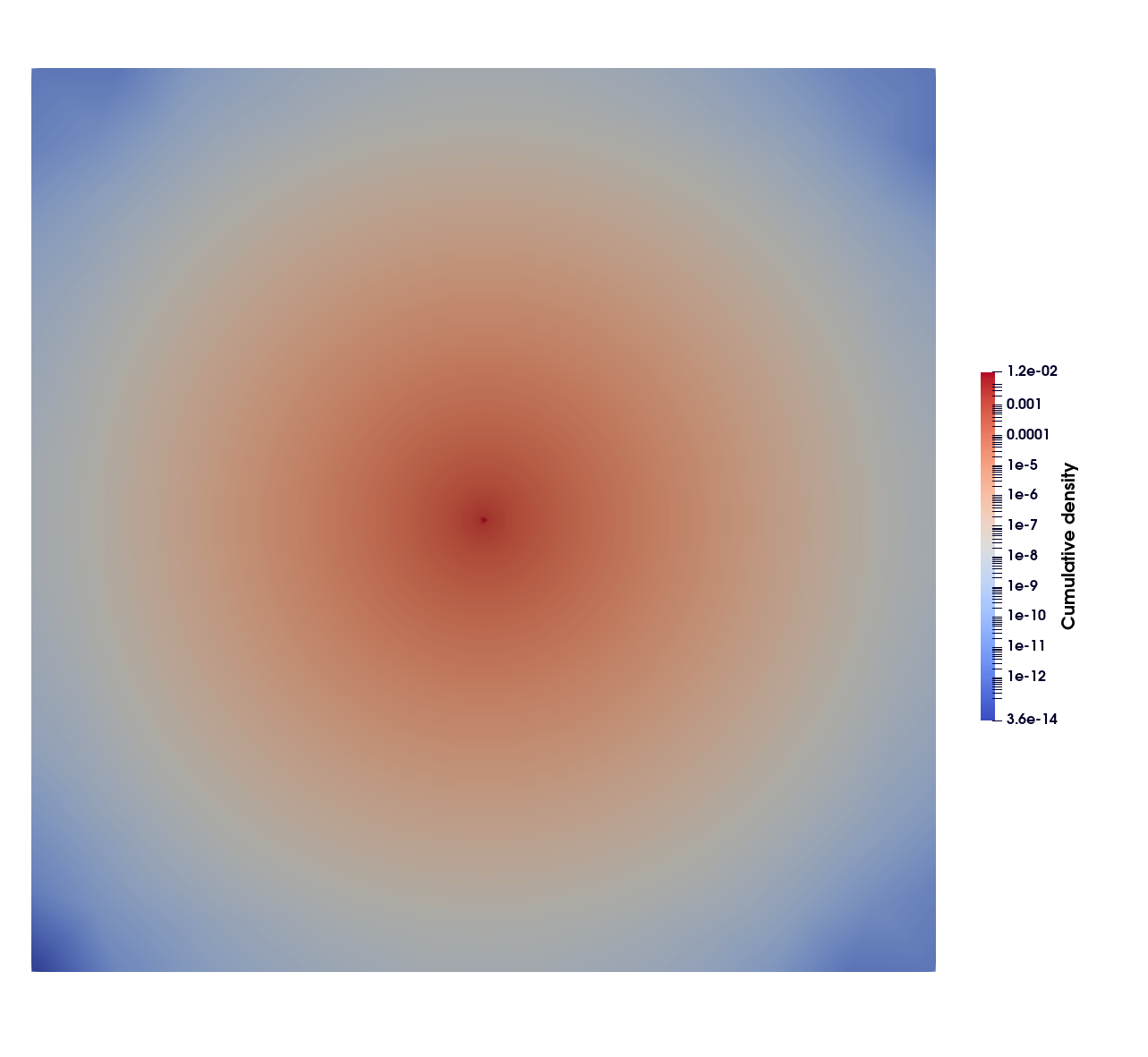}}
    \subfigure[]{\includegraphics[width = 0.49\textwidth]{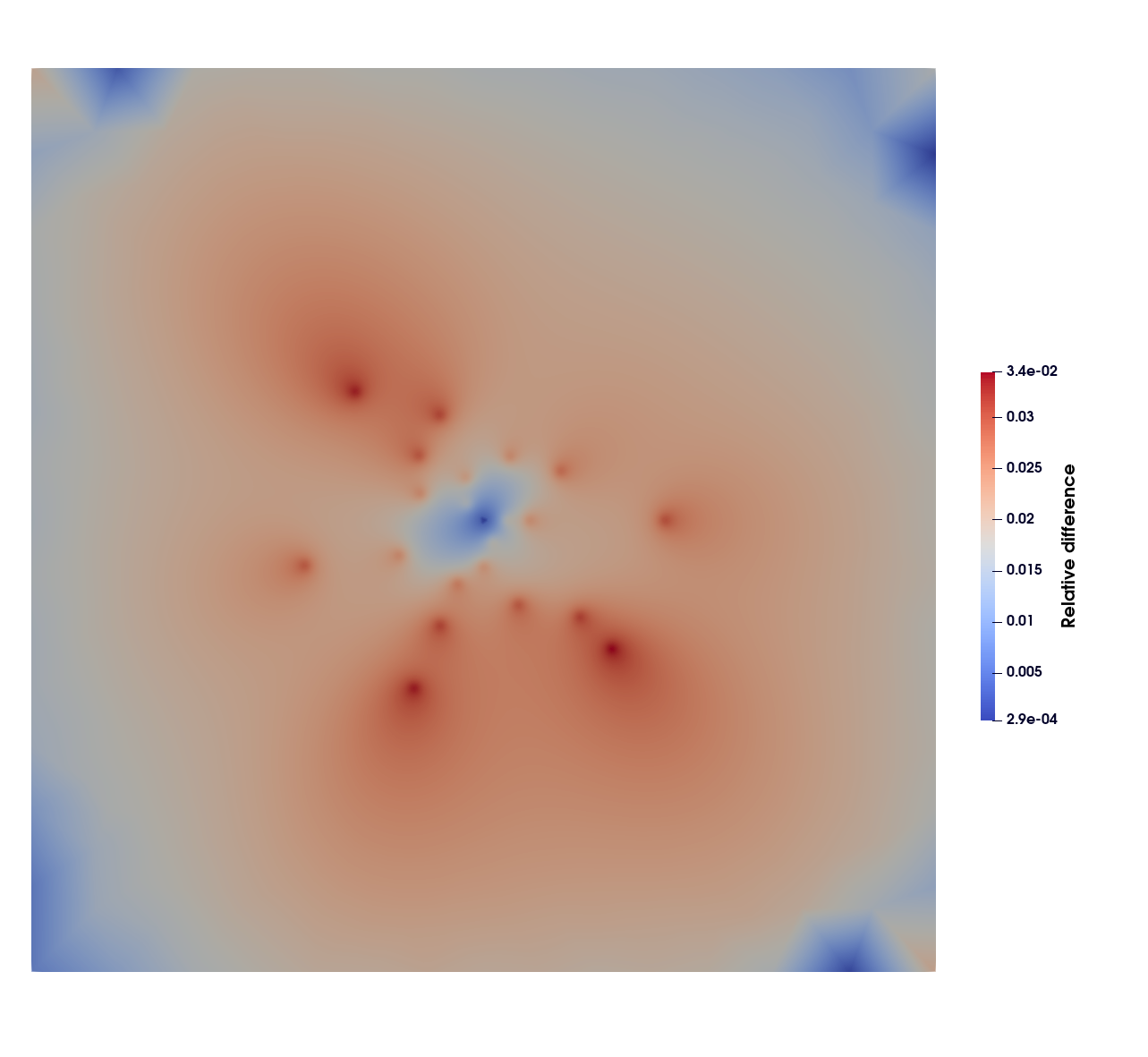}}
	\caption{(a) Cumulative mosquito population density $c(x)= \int_0^{20} h(s,x)\,ds$ obtained by solving the reaction–diffusion model \eqref{eq:EDP_pop_dens} with a homogeneous diffusion coefficient and the MLE parameter vector $\Theta^*$. 
 (b) Relative difference $(c_n(x)-c(x))/c_n(x)$ between the cumulative density with trapping (parameters $\Theta^*$) and without trapping ($\gamma=0$, with $\sigma$ and $\nu$ fixed at their MLE values), highlighting the localized impact of traps.}
	\label{fig:simus2D}
\end{figure}

With the parameter vector $\Theta^*$, we simulated the unobserved mosquito dynamics using the reaction–diffusion model \eqref{eq:EDP_pop_dens}. Figure~\ref{fig:simus2D}a displays the cumulative density $c(x)=\int_0^{20} h(s,x)\,ds$. Because the trapping effect is not immediately apparent on $c(x)$, Fig.~\ref{fig:simus2D}b shows the relative difference $(c_n(x)-c(x))/c_n(x)$, where $c_n(x)=\int_0^{20} h_n(s,x)\,ds$ and $h_n$ solves \eqref{eq:EDP_pop_dens} with $\gamma=0$ (keeping $\sigma$ and $\nu$ at their MLEs, given by $\Theta^*$). The reduction near traps confirms the expected local impact of trapping.

\begin{figure}[h!]
\centering
\includegraphics[width=\textwidth]{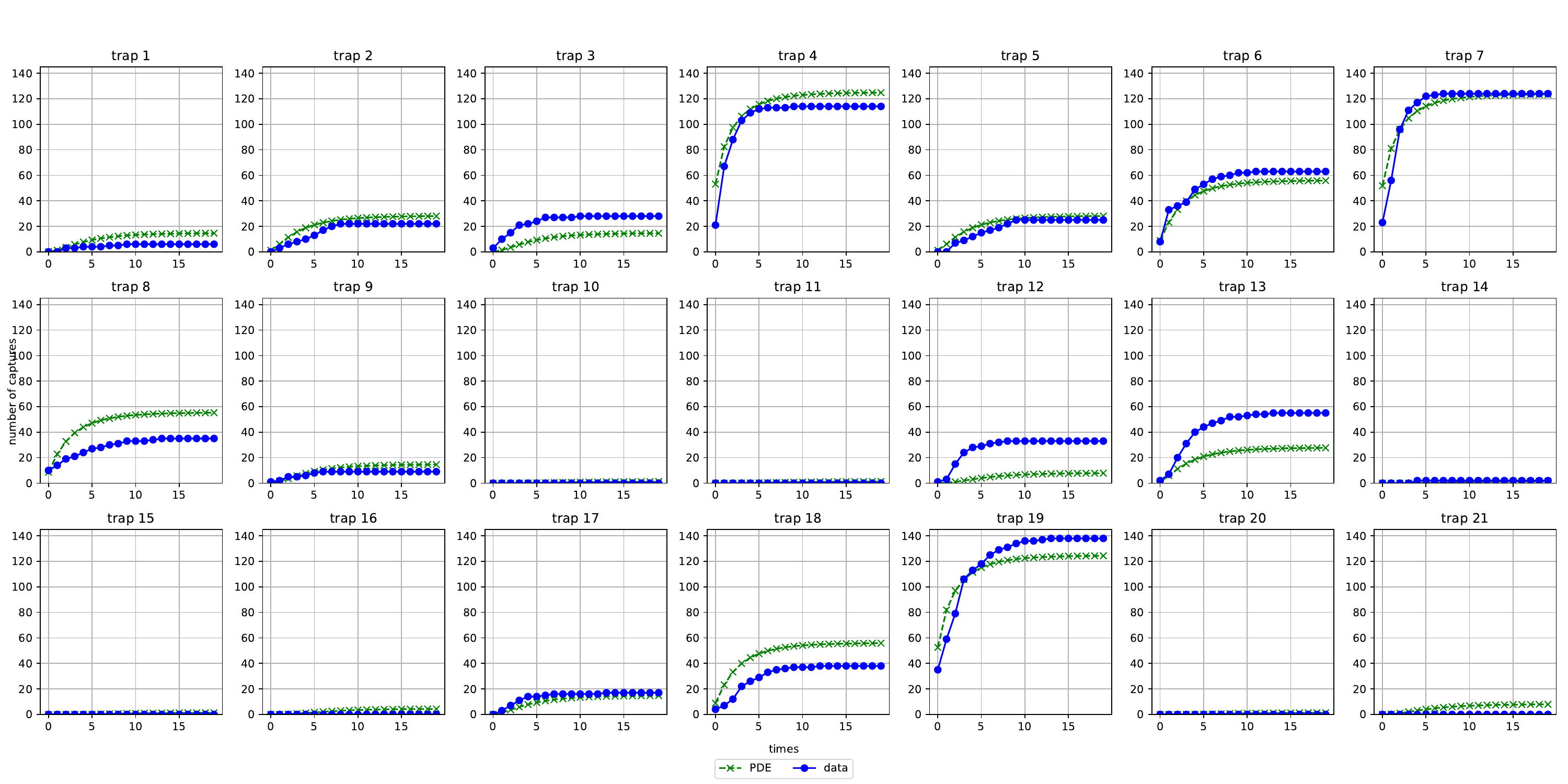}
\caption{Observed trap counts (blue circles) versus expected captures (green crosses) predicted by \eqref{eqn:pi}–\eqref{eq:EDP_pop_dens} under the homogeneous model with MLE parameters $\Theta^*$. Each panel corresponds to one trap (1–21). All panels share the same axes (days on the $x$-axis; cumulative captures on the $y$-axis), enabling direct comparison across traps.}
\label{fig:DataVsPde}
\end{figure}

In Fig.~\ref{fig:DataVsPde}, we compare the expected number of trapped individuals predicted by \eqref{eqn:pi}–\eqref{eq:EDP_pop_dens} with parameter vector $\Theta^*$ (homogeneous case) to the observed counts. The model reproduces the rapid early accumulation (days 0--5) and subsequent saturation of cumulative captures across most traps, as well as the near-zero captures at several peripheral traps. Some discrepancies remain at a subset of sites: the model tends to overpredict  some central traps (e.g., \#8, \#18)  and underpredict some intermerdiate traps (e.g., \#12, \#13). These residual patterns are consistent with unmodeled trap-level heterogeneity (e.g., variability in $\gamma$ or local micro-environmental effects).

\section{Discussion}

\paragraph{Summary of contributions.}
In the present work, we unveiled the dynamics of mosquito populations in mark–release–recapture (MRR) experiments using coupled mechanistic individual- and population-level frameworks. At the microscopic scale, we formulated an individual-based model (IBM) that encodes movement, mortality, and capture for each mosquito. At the macroscopic scale, we derived a reaction–diffusion partial differential equation (PDE) that is the deterministic limit of the IBM and yields, for any parameter vector, the expected population density and capture intensities at each trap without simulating the IBM explicitly. Embedding this PDE model in a mechanistic–statistical observation scheme allowed us to estimate movement, post-release survival, and trapping rate despite the severe information loss inherent to MRR data. A simulation study validated the procedure and showed accurate estimation of the dispersal, survival, and trapping parameters. We further examined spatial heterogeneity by contrasting a homogeneous mobility field with a heterogeneous one (urban versus non-residential areas) and found that introducing heterogeneity did not significantly improve the fit to our data; parsimony therefore favored the homogeneous model.

\paragraph{Strengths of a mechanistic approach and comparison with previous work.}
Classical analyses of MRR datasets often summarize dispersal and survival empirically from trap returns  \citep[e.g., exponential fits for daily survival, recapture-weighted averages for dispersal,][]{MuiKay98}. In contrast, our approach links movement, mortality, and capture through explicit mechanisms and computes likelihoods from model-predicted captures at each trap and day, which avoids ad hoc transformations and, crucially, clarifies the interpretation of dispersal metrics. Our estimates are consistent with those of \cite{GAT21} for lifespan: in that other Cuban field trial, the probability of daily survival (PDS) was obtained by fitting an exponential model to recapture counts, and the average post-release life expectancy was computed as $1/(-\ln\mathrm{PDS})$, yielding 3.76~days; combined with the $\sim$3-day age at release, this corresponds to a total lifespan close to 7~days, in good agreement with our inference (about 8~days). For context, published field MRR studies of \emph{wild} (non-sterile) \emph{A. aegypti} males report daily survival probabilities around $0.57$–$0.70$ in northern Australia \citep{MuiKay98} (implying mean life expectancy of $\approx 1.8$–$2.8$~days) and $\approx 0.77$ on the Kenyan coast \citep{McD77I} (implying $\approx 3.8$~days). Regarding pre-release age, \cite{MuiKay98} report that released males were 1–2 days post-emergence, whereas \cite{McD77I} describes ``newly emerged" males without specifying an exact number of days. These values place our sterile-male post-release estimate ($5$~days) and total lifespan (8~days) slightly above these wild-male estimates.

By contrast, dispersal metrics differ substantially. In  \citep{GAT21}, dispersal was summarized from the locations of recaptured marked males in a BG-Sentinel network arranged in concentric rings up to 400~m, as a mean distance of recaptures (77.3~m) and as flight ranges (43.2~m for 50\% of recaptures, 110.5~m for 90\%). Methodologically, this “mean distance” is the average of trap-to-release distances weighted by how many mosquitoes were recaptured in each trap; in other words, it is the mean distance conditional on being captured (this is a standard approach, also used in \citep{McD77II,MuiKay98}). Such a quantity is generally a downward-biased proxy for the population mean displacement because (i) capture probability typically declines with distance as trap density per unit circumference decreases with increasing ring radius, although this effect can be corrected for by applying a weighting factor, as proposed by \cite{MuiKay98}; (ii) long-distance movements are right-truncated by the outer trap radius (400~m); and (iii) capture is an absorbing event that truncates trajectories—individuals caught early cannot contribute longer displacements they might otherwise have achieved. For these reasons, the reported distances should be interpreted as lower bounds on true dispersal. In our mechanistic–statistical framework, dispersal is inferred through an explicit diffusion parameter ($\sigma$), independent of trap placement. With the homogeneous-model estimate $\sigma=64~\text{m}/\sqrt{\text{day}}$, the expected distance from the release point is $E(\|X_t\|)=\sigma\sqrt{\pi t/2}\approx 80\sqrt{t}$~m, i.e., roughly 180~m after 5~days, naturally exceeding the uncorrected mean of recaptures and aligning with the expectation that trap-weighted averages under-estimate long-range movement.

\paragraph{Computational considerations relative to IBM-based inference.}
An additional advantage of our framework emerges when comparing the PDE-based inference to a purely IBM-based estimation strategy. Because the IBM is stochastic, evaluating the likelihood would require simulating a large number of independent trajectories at each candidate parameter vector to stabilize Monte Carlo noise. This quickly becomes computationally prohibitive. By contrast, the PDE-based likelihood is deterministic: optimization requires only forward PDE simulations. We therefore reserved the IBM for tasks where stochasticity is essential, namely (i) simulation-based validation of the inference procedure and (ii) parametric bootstrap to propagate process and observation noise into standard errors.

\paragraph{Identifiability, parsimony, and heterogeneity.}
Our likelihood analysis shows that allowing two mobility coefficients (urban vs.\ non-residential) raises the maximum likelihood but does not improve AIC relative to the homogeneous model once penalizing the additional parameter. Biologically, capture counts in our design are explained by an effective, spatially averaged mobility; this does not rule out genuine heterogeneity but suggests its imprint on daily indoor trap counts is weak compared with mortality and capture processes at the scales considered.

\paragraph{Experimental design and practical implications.}
Operationally, the diffusion estimate indicates a typical sterile male is expected to be $\approx 180$~m from the release point after 5~days, consistent with SIT release grids on the order of a few hundred meters. Our trap network was informative across radii but thinned at the periphery: no captures were recorded in traps \#16 and \#21 (at 250 and 300~m), and in traps \#10, \#11, \#15, \#20 (at 400~m), while all other traps registered at least one capture. 

Regarding potential improvements to the MRR experiment, a promising alternative would be to perform simultaneous releases of mosquitoes marked with distinct colors at different locations, rather than successive weekly releases from the same site. In such a design, frequency data (i.e., the relative proportions of each color captured per trap) could be analyzed instead of absolute abundances, as proposed in \citep{RoqWal16}. Using proportions has the advantage of reducing sensitivity to heterogeneity in trapping efficiency across sites (e.g., spatial variations in $\gamma$ or $R$), thereby mitigating potential biases in parameter estimation.

\paragraph{Limitations and future work.}
Our analysis did not account for environmental or ecological covariates such as weather conditions or host density, although these factors likely influence both movement and capture. The Gaussian trap kernel with a fixed $R$ is also a simplification. Moreover, the Brownian framework used here represents an \emph{effective} mobility that combines periods of active flight and resting; as a result, the estimated $\sigma$ incorporates pausing behavior, and true flight speeds during active bouts are expected to be higher than those implied by $\sigma$ alone. Future work will aim to extend the framework to include time-varying mobility and mortality, alternative trap kernels, and explicit environmental drivers (e.g., wind, temperature, presence of hosts).
Beyond this specific case, the methodology is transferable to other species and experimental designs, and it naturally accommodates richer setups (e.g., the above-mentioned simultaneous multi-source releases) that can further enhance identifiability and robustness.

\section*{Acknowledgments} 
This work received support from the French government, managed by the National Research Agency (ANR), under the France 2030 program, reference ANR 23 EXMA 0001. In particular, this project has received financial support from the PEPR Maths-Vives (through a Chaire Internationale that made possible a 3-month stay of René Gato in Paris and the project Maths-ArboV under grant agreement ANR-24-EXMA-0004). It also benefited from the  \textit{Scientific stays for PhD students} program of the Fondation Sciences Mathématiques de Paris for N. Nguyen's two-month stay at INRAE.

\bibliographystyle{chicago}
%\bibliography{Meca-stat}

\clearpage

\section*{Appendices}

\section*{Appendix 1: Datasets}

\begin{table}[h!]
\centering
\caption{Trap Data 1.}
\scriptsize % Reduces the font size
\begin{tabular}{|c|cccccccccccccccccccc|}
\hline
Day &  0 &  1 &  2 &  3 &  4 &  5 &  6 &  7 &  8 &  9 &  10 &  11 &  12 &  13 &  14 &  15 &  16 &  17 &  18 &  19 \\
\hline
Trap 1  &      0 &      0 &      0 &      0 &      0 &      0 &      0 &      1 &      0 &      1 &       0 &       0 &       0 &       0 &       0 &       0 &       0 &       0 &       0 &       0 \\
Trap 2  &      0 &      1 &      1 &      2 &      1 &      2 &      2 &      0 &      0 &      0 &       0 &       0 &       0 &       0 &       0 &       0 &       0 &       0 &       0 &       0 \\
Trap 3  &      0 &      1 &      1 &      1 &      0 &      0 &      0 &      0 &      0 &      0 &       0 &       0 &       0 &       0 &       0 &       0 &       0 &       0 &       0 &       0 \\
Trap 4  &      5 &      5 &      3 &      3 &      0 &      0 &      1 &      0 &      0 &      0 &       0 &       0 &       0 &       0 &       0 &       0 &       0 &       0 &       0 &       0 \\
Trap 5  &      0 &      0 &      0 &      1 &      0 &      2 &      1 &      0 &      0 &      2 &       0 &       0 &       0 &       0 &       0 &       0 &       0 &       0 &       0 &       0 \\
Trap 6  &      3 &      2 &      2 &      1 &      6 &      0 &      0 &      1 &      1 &      0 &       0 &       0 &       0 &       0 &       0 &       0 &       0 &       0 &       0 &       0 \\
Trap 7  &     13 &     20 &      4 &      6 &      3 &      3 &      1 &      1 &      0 &      0 &       0 &       0 &       0 &       0 &       0 &       0 &       0 &       0 &       0 &       0 \\
Trap 8  &      0 &      0 &      1 &      0 &      0 &      0 &      0 &      0 &      0 &      0 &       0 &       0 &       0 &       0 &       0 &       0 &       0 &       0 &       0 &       0 \\
Trap 9  &      0 &      0 &      0 &      0 &      1 &      0 &      1 &      0 &      0 &      0 &       0 &       0 &       0 &       0 &       0 &       0 &       0 &       0 &       0 &       0 \\
Trap 10 &      0 &      0 &      0 &      0 &      0 &      0 &      0 &      0 &      0 &      0 &       0 &       0 &       0 &       0 &       0 &       0 &       0 &       0 &       0 &       0 \\
Trap 11 &      0 &      0 &      0 &      0 &      0 &      0 &      0 &      0 &      0 &      0 &       0 &       0 &       0 &       0 &       0 &       0 &       0 &       0 &       0 &       0 \\
Trap 12 &      0 &      2 &      7 &      0 &      0 &      0 &      0 &      1 &      1 &      0 &       0 &       0 &       0 &       0 &       0 &       0 &       0 &       0 &       0 &       0 \\
Trap 13 &      0 &      1 &      2 &      3 &      2 &      1 &      1 &      0 &      0 &      0 &       0 &       0 &       0 &       0 &       0 &       0 &       0 &       0 &       0 &       0 \\
Trap 14 &      0 &      0 &      0 &      0 &      0 &      0 &      0 &      0 &      0 &      0 &       0 &       0 &       0 &       0 &       0 &       0 &       0 &       0 &       0 &       0 \\
Trap 15 &      0 &      0 &      0 &      0 &      0 &      0 &      0 &      0 &      0 &      0 &       0 &       0 &       0 &       0 &       0 &       0 &       0 &       0 &       0 &       0 \\
Trap 16 &      0 &      0 &      0 &      0 &      0 &      0 &      0 &      0 &      0 &      0 &       0 &       0 &       0 &       0 &       0 &       0 &       0 &       0 &       0 &       0 \\
Trap 17 &      0 &      3 &      2 &      0 &      1 &      0 &      1 &      0 &      0 &      0 &       0 &       0 &       0 &       0 &       0 &       0 &       0 &       0 &       0 &       0 \\
Trap 18 &      0 &      0 &      5 &      2 &      1 &      1 &      1 &      1 &      1 &      1 &       0 &       0 &       0 &       0 &       0 &       0 &       0 &       0 &       0 &       0 \\
Trap 19 &     21 &      5 &      4 &      2 &      5 &      2 &      3 &      1 &      0 &      1 &       0 &       0 &       0 &       0 &       0 &       0 &       0 &       0 &       0 &       0 \\
Trap 20 &      0 &      0 &      0 &      0 &      0 &      0 &      0 &      0 &      0 &      0 &       0 &       0 &       0 &       0 &       0 &       0 &       0 &       0 &       0 &       0 \\
Trap 21 &      0 &      0 &      0 &      0 &      0 &      0 &      0 &      0 &      0 &      0 &       0 &       0 &       0 &       0 &       0 &       0 &       0 &       0 &       0 &       0 \\
\hline
\end{tabular}
\end{table}

\begin{table}[ht]
\centering
\caption{Trap Data 2.}
\scriptsize % Reduces the font size
\begin{tabular}{|c|cccccccccccccccccccc|}
\hline
Day &  0 &  1 &  2 &  3 &  4 &  5 &  6 &  7 &  8 &  9 &  10 &  11 &  12 &  13 &  14 &  15 &  16 &  17 &  18 &  19 \\ 
\hline
Trap 1  &  0 &  0 &  0 &  0 &  0 &  0 &  0 &  0 &  0 &  0 &  0 &  0 &  0 &  0 &  0 &  0 &  0 &  0 &  0 &  0 \\
Trap 2  &  0 &  1 &  2 &  0 &  1 &  0 &  0 &  2 &  2 &  0 &  0 &  0 &  0 &  0 &  0 &  0 &  0 &  0 &  0 &  0 \\
Trap 3  &  2 &  0 &  3 &  0 &  0 &  0 &  1 &  0 &  0 &  0 &  1 &  0 &  0 &  0 &  0 &  0 &  0 &  0 &  0 &  0 \\
Trap 4  &  6 & 22 &  7 &  1 &  2 &  1 &  0 &  0 &  0 &  0 &  0 &  0 &  0 &  0 &  0 &  0 &  0 &  0 &  0 &  0 \\
Trap 5  &  0 &  0 &  1 &  1 &  1 &  1 &  1 &  0 &  0 &  0 &  0 &  0 &  0 &  0 &  0 &  0 &  0 &  0 &  0 &  0 \\
Trap 6  &  2 &  1 &  1 &  1 &  1 &  1 &  1 &  0 &  0 &  0 &  0 &  1 &  0 &  0 &  0 &  0 &  0 &  0 &  0 &  0 \\
Trap 7  &  3 &  4 &  3 &  3 &  0 &  0 &  0 &  0 &  0 &  0 &  0 &  0 &  0 &  0 &  0 &  0 &  0 &  0 &  0 &  0 \\
Trap 8  &  4 &  0 &  2 &  2 &  0 &  0 &  0 &  0 &  1 &  1 &  0 &  0 &  1 &  1 &  0 &  0 &  0 &  0 &  0 &  0 \\
Trap 9  &  0 &  0 &  1 &  0 &  0 &  1 &  0 &  0 &  0 &  0 &  0 &  0 &  0 &  0 &  0 &  0 &  0 &  0 &  0 &  0 \\
Trap 10 &  0 &  0 &  0 &  0 &  0 &  0 &  0 &  0 &  0 &  0 &  0 &  0 &  0 &  0 &  0 &  0 &  0 &  0 &  0 &  0 \\
Trap 11 &  0 &  0 &  0 &  0 &  0 &  0 &  0 &  0 &  0 &  0 &  0 &  0 &  0 &  0 &  0 &  0 &  0 &  0 &  0 &  0 \\
Trap 12 &  0 &  0 &  3 &  0 &  0 &  0 &  1 &  0 &  0 &  0 &  0 &  0 &  0 &  0 &  0 &  0 &  0 &  0 &  0 &  0 \\
Trap 13 &  1 &  0 &  8 &  3 &  2 &  1 &  1 &  1 &  2 &  0 &  1 &  1 &  0 &  1 &  0 &  0 &  0 &  0 &  0 &  0 \\
Trap 14 &  0 &  0 &  0 &  0 &  0 &  0 &  0 &  0 &  0 &  0 &  0 &  0 &  0 &  0 &  0 &  0 &  0 &  0 &  0 &  0 \\
Trap 15 &  0 &  0 &  0 &  0 &  0 &  0 &  0 &  0 &  0 &  0 &  0 &  0 &  0 &  0 &  0 &  0 &  0 &  0 &  0 &  0 \\
Trap 16 &  0 &  0 &  0 &  0 &  0 &  0 &  0 &  0 &  0 &  0 &  0 &  0 &  0 &  0 &  0 &  0 &  0 &  0 &  0 &  0 \\
Trap 17 &  0 &  0 &  0 &  1 &  1 &  0 &  0 &  0 &  0 &  0 &  0 &  0 &  0 &  1 &  0 &  0 &  0 &  0 &  0 &  0 \\
Trap 18 &  3 &  1 &  0 &  1 &  2 &  0 &  1 &  1 &  0 &  0 &  0 &  0 &  1 &  0 &  0 &  0 &  0 &  0 &  0 &  0 \\
Trap 19 &  8 & 12 &  5 &  3 &  0 &  0 &  1 &  0 &  0 &  0 &  0 &  0 &  0 &  1 &  0 &  0 &  0 &  0 &  0 &  0 \\
Trap 20 &  0 &  0 &  0 &  0 &  0 &  0 &  0 &  0 &  0 &  0 &  0 &  0 &  0 &  0 &  0 &  0 &  0 &  0 &  0 &  0 \\
Trap 21 &  0 &  0 &  0 &  0 &  0 &  0 &  0 &  0 &  0 &  0 &  0 &  0 &  0 &  0 &  0 &  0 &  0 &  0 &  0 &  0 \\ 
\hline
\end{tabular}
\end{table}

\begin{table}[ht]
\centering
\caption{Trap Data 3.}
\scriptsize % Reduces the font size
\begin{tabular}{|c|cccccccccccccccccccc|}
\hline
Day &  0 &  1 &  2 &  3 &  4 &  5 &  6 &  7 &  8 &  9 &  10 &  11 &  12 &  13 &  14 &  15 &  16 &  17 &  18 &  19 \\
\hline
Trap 1 &  0 &   0 &   0 &   0 &   1 &   0 &   0 &   0 &   0 &   0 &   0 &   0 &   0 &   0 &   0 &   0 &   0 &   0 & 
   0 &   0  \\
 Trap 2 &  0 &   0 &   0 &   0 &   0 &   0 &   1 &   0 &   0 &   0 &   0 &   0 &   0 &   0 &   0 &   0 &   0 &   0 & 
   0 &   0  \\
 Trap 3 &  0 &   1 &   1 &   3 &   0 &   1 &   2 &   0 &   0 &   0 &   0 &   0 &   0 &   0 &   0 &   0 &   0 &   0 & 
   0 &   0  \\
 Trap 4 & 10 &   1 &   3 &   6 &   3 &   0 &   0 &   0 &   0 &   0 &   0 &   0 &   0 &   0 &   0 &   0 &   0 &   0 & 
   0 &   0  \\
 Trap 5 &  0 &   0 &   0 &   0 &   0 &   0 &   0 &   0 &   2 &   1 &   0 &   0 &   0 &   0 &   0 &   0 &   0 &   0 & 
   0 &   0  \\
 Trap 6  &  1 &  10 &   0 &   1 &   1 &   2 &   2 &   0 &   0 &   2 &   0 &   0 &   0 &   0 &   0 &   0 &   0 &   0 & 
   0 &   0  \\
 Trap 7 &  7 &   0 &   1 &   2 &   0 &   0 &   0 &   0 &   0 &   0 &   0 &   0 &   0 &   0 &   0 &   0 &   0 &   0 & 
   0 &   0  \\
 Trap 8 &  4 &   3 &   0 &   0 &   2 &   2 &   1 &   1 &   0 &   0 &   0 &   0 &   0 &   0 &   0 &   0 &   0 &   0 & 
   0 &   0  \\
 Trap 9 &  0 &   0 &   2 &   0 &   0 &   1 &   0 &   0 &   0 &   0 &   0 &   0 &   0 &   0 &   0 &   0 &   0 &   0 & 
   0 &   0  \\
 Trap 10 &  0 &   0 &   0 &   0 &   0 &   0 &   0 &   0 &   0 &   0 &   0 &   0 &   0 &   0 &   0 &   0 &   0 &   0 & 
   0 &   0  \\
 Trap 11 &  0 &   0 &   0 &   0 &   0 &   0 &   0 &   0 &   0 &   0 &   0 &   0 &   0 &   0 &   0 &   0 &   0 &   0 & 
   0 &   0  \\
 Trap 12 &  0 &   0 &   0 &   7 &   2 &   1 &   1 &   0 &   0 &   0 &   0 &   0 &   0 &   0 &   0 &   0 &   0 &   0 & 
   0 &   0  \\
 Trap 13 &  0 &   4 &   0 &   0 &   0 &   1 &   0 &   0 &   1 &   0 &   0 &   0 &   0 &   0 &   0 &   0 &   0 &   0 & 
   0 &   0  \\
 Trap 14 &  0 &   0 &   0 &   0 &   0 &   0 &   0 &   0 &   0 &   0 &   0 &   0 &   0 &   0 &   0 &   0 &   0 &   0 & 
   0 &   0  \\
 Trap 15 &  0 &   0 &   0 &   0 &   0 &   0 &   0 &   0 &   0 &   0 &   0 &   0 &   0 &   0 &   0 &   0 &   0 &   0 & 
   0 &   0  \\
 Trap 16 &  0 &   0 &   0 &   0 &   0 &   0 &   0 &   0 &   0 &   0 &   0 &   0 &   0 &   0 &   0 &   0 &   0 &   0 & 
   0 &   0  \\
 Trap 17 &  0 &   0 &   0 &   3 &   1 &   0 &   0 &   1 &   0 &   0 &   0 &   0 &   0 &   0 &   0 &   0 &   0 &   0 & 
   0 &   0  \\
 Trap 18 &  0 &   0 &   0 &   2 &   0 &   1 &   1 &   0 &   0 &   0 &   0 &   0 &   0 &   0 &   0 &   0 &   0 &   0 & 
   0 &   0  \\
 Trap 19 &  4 &   1 &   1 &   1 &   1 &   2 &   3 &   0 &   0 &   0 &   0 &   0 &   0 &   0 &   0 &   0 &   0 &   0 & 
   0 &   0  \\
 Trap 20 &  0 &   0 &   0 &   0 &   0 &   0 &   0 &   0 &   0 &   0 &   0 &   0 &   0 &   0 &   0 &   0 &   0 &   0 & 
   0 &   0  \\
 Trap 21 &  0 &   0 &   0 &   0 &   0 &   0 &   0 &   0 &   0 &   0 &   0 &   0 &   0 &   0 &   0 &   0 &   0 &   0 & 
   0 &   0  \\
\hline
\end{tabular}
\end{table}

\begin{table}[ht]
\centering
\caption{Trap Data 4.}
\scriptsize % Reduces the font size
\begin{tabular}{|c|cccccccccccccccccccc|}
\hline
Day &  0 &  1 &  2 &  3 &  4 &  5 &  6 &  7 &  8 &  9 &  10 &  11 &  12 &  13 &  14 &  15 &  16 &  17 &  18 &  19 \\
\hline
Trap 1 & 0 &   0 &   3 &   0 &   0 &   0 &   0 &   0 &   0 &   0 &   0 &   0 &   0 &   0 &   0 &   0 &   0 &   0 & 
   0 &   0 \\
 Trap 2 & 0 &   1 &   0 &   0 &   0 &   1 &   1 &   1 &   0 &   0 &   0 &   0 &   0 &   0 &   0 &   0 &   0 &   0 & 
   0 &   0 \\
 Trap 3 & 1 &   5 &   0 &   2 &   1 &   1 &   0 &   0 &   0 &   0 &   0 &   0 &   0 &   0 &   0 &   0 &   0 &   0 & 
   0 &   0 \\
 Trap 4 & 0 &  18 &   8 &   5 &   1 &   2 &   0 &   0 &   0 &   1 &   0 &   0 &   0 &   0 &   0 &   0 &   0 &   0 & 
   0 &   0 \\
 Trap 5 & 0 &   0 &   6 &   0 &   2 &   0 &   0 &   2 &   1 &   0 &   0 &   0 &   0 &   0 &   0 &   0 &   0 &   0 & 
   0 &   0 \\
 Trap 6 & 2 &  12 &   0 &   0 &   2 &   1 &   1 &   1 &   0 &   0 &   0 &   0 &   0 &   0 &   0 &   0 &   0 &   0 & 
   0 &   0 \\
 Trap 7 & 0 &   9 &  32 &   4 &   3 &   2 &   0 &   0 &   0 &   0 &   0 &   0 &   0 &   0 &   0 &   0 &   0 &   0 & 
   0 &   0 \\
 Trap 8 & 2 &   1 &   2 &   0 &   1 &   1 &   0 &   1 &   0 &   1 &   0 &   0 &   0 &   0 &   0 &   0 &   0 &   0 & 
   0 &   0 \\
 Trap 9 & 1 &   1 &   0 &   0 &   0 &   0 &   0 &   0 &   0 &   0 &   0 &   0 &   0 &   0 &   0 &   0 &   0 &   0 & 
   0 &   0 \\
 Trap 10 & 0 &   0 &   0 &   0 &   0 &   0 &   0 &   0 &   0 &   0 &   0 &   0 &   0 &   0 &   0 &   0 &   0 &   0 & 
   0 &   0 \\
 Trap 11 & 0 &   0 &   0 &   0 &   0 &   0 &   0 &   0 &   0 &   0 &   0 &   0 &   0 &   0 &   0 &   0 &   0 &   0 & 
   0 &   0 \\
 Trap 12 & 1 &   0 &   2 &   2 &   2 &   0 &   0 &   0 &   0 &   0 &   0 &   0 &   0 &   0 &   0 &   0 &   0 &   0 & 
   0 &   0 \\
 Trap 13 & 1 &   0 &   3 &   5 &   5 &   1 &   1 &   1 &   0 &   0 &   0 &   0 &   0 &   0 &   0 &   0 &   0 &   0 & 
   0 &   0 \\
 Trap 14 & 0 &   0 &   0 &   0 &   2 &   0 &   0 &   0 &   0 &   0 &   0 &   0 &   0 &   0 &   0 &   0 &   0 &   0 & 
   0 &   0 \\
 Trap 15 & 0 &   0 &   0 &   0 &   0 &   0 &   0 &   0 &   0 &   0 &   0 &   0 &   0 &   0 &   0 &   0 &   0 &   0 & 
   0 &   0 \\
 Trap 16 & 0 &   0 &   0 &   0 &   0 &   0 &   0 &   0 &   0 &   0 &   0 &   0 &   0 &   0 &   0 &   0 &   0 &   0 & 
   0 &   0 \\
 Trap 17 & 0 &   0 &   2 &   0 &   0 &   0 &   0 &   0 &   0 &   0 &   0 &   0 &   0 &   0 &   0 &   0 &   0 &   0 & 
   0 &   0 \\
 Trap 18 & 1 &   2 &   0 &   5 &   1 &   1 &   1 &   0 &   0 &   0 &   0 &   0 &   0 &   0 &   0 &   0 &   0 &   0 & 
   0 &   0 \\
 Trap 19 & 2 &   6 &  10 &  21 &   1 &   1 &   0 &   3 &   2 &   2 &   2 &   0 &   1 &   0 &   0 &   0 &   0 &   0 & 
   0 &   0 \\
 Trap 20 & 0 &   0 &   0 &   0 &   0 &   0 &   0 &   0 &   0 &   0 &   0 &   0 &   0 &   0 &   0 &   0 &   0 &   0 & 
   0 &   0 \\
 Trap 21 & 0 &   0 &   0 &   0 &   0 &   0 &   0 &   0 &   0 &   0 &   0 &   0 &   0 &   0 &   0 &   0 &   0 &   0 & 
   0 &   0 \\
\hline
\end{tabular}
\end{table}

\clearpage

\section*{Appendix 2: PDE vs microscopic model, 1D case}

	\begin{figure}
		\centering
		\includegraphics[width = 1\textwidth]{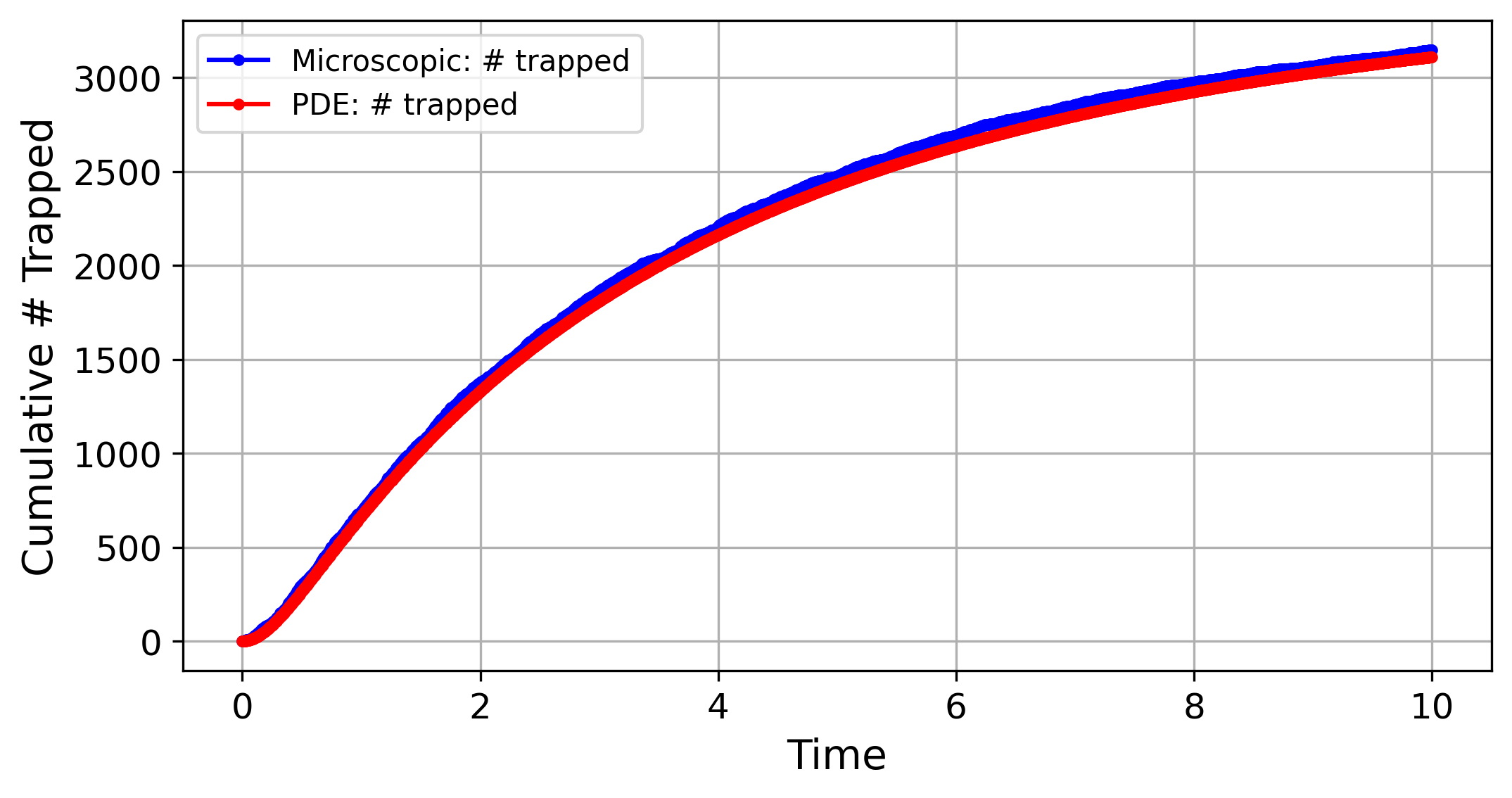}
            \caption{Expected number of trapped mosquitoes predicted by the macroscopic model \eqref{eq:EDP_pop_dens} in one dimension, expressed as $N_0 \, \pi_1(t) = \int_0^t \int_{\R} f_1(s) \, h(s,x) \, dx \, ds$, compared with the actual number of trapped mosquitoes obtained from one simulation of the microscopic model. The assumptions are the same as in Fig.~\ref{fig:1D}.}
 
		\label{fig:trap1D}
	\end{figure}

\clearpage

\section*{Appendix 3: PDE mesh}

\begin{figure}[h!]
		\centering
		\includegraphics[width = \textwidth]{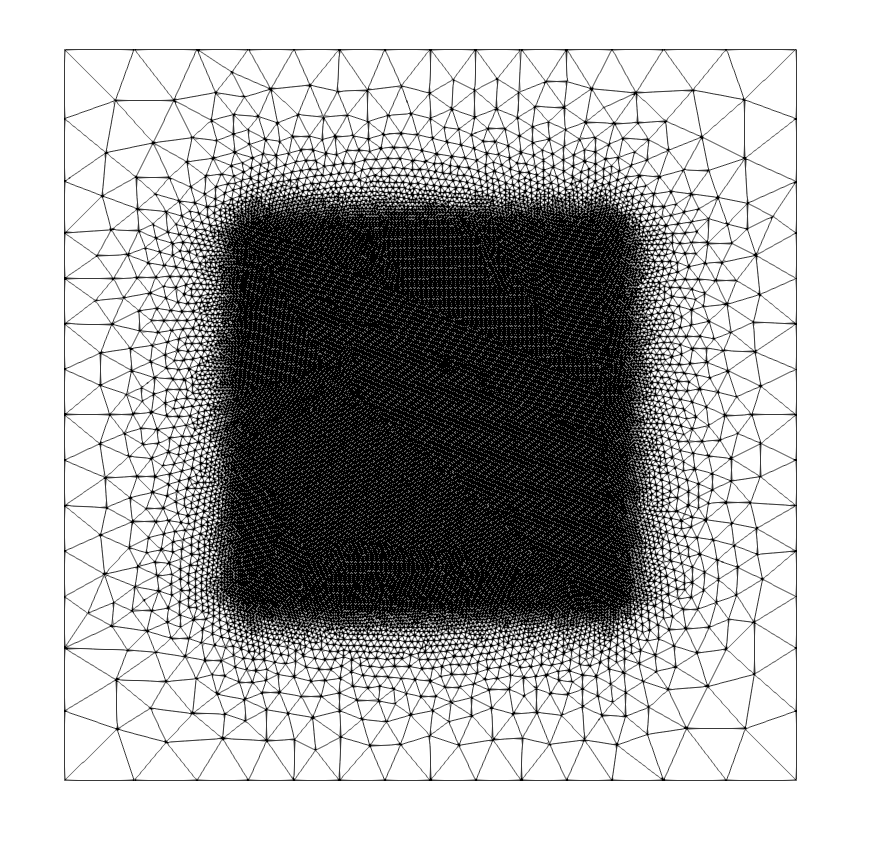}
		\caption{The mesh was chosen so that further reductions in mesh size do not affect the simulation results. It was refined in the regions where trapping occurs and consists of 32197 nodes.}
		\label{fig:mesh}
	\end{figure}

\section*{Appendix~4: Algorithmic description of the generation of the simulated data}
Using the microscopic model developed in Section~\ref{subsec:models}, we generated datasets that mimic the real data and check if the parameters are  correctly identified by our inference procedure. We first fixed a parameter $\Theta_{\text{true}}$ and used the individual-based model built in Section~\ref{subsec:models} to generate simulated data $\{\tilde{y}_i^j, \ i=1,\ldots,21, \ j=0,\ldots,19\}$ by the following steps:
	\begin{itemize}
		\item We create the maps of landscapes and positions of $n_P = 21$ traps $\omega_i = B_R(q_i)$, with $i = 1, \dots, 21$. We fix $R = 10$m. 
		\item We generate $N_0 = 10^4$ stochastic processes $X_t^k$ with $k = 1, \dots, N_0$, by an Euler--Maruyama discretization of  \eqref{eqn:process}, and the release point $x_0 = (0,0)$. 
		\item For each individual, we generate its lifetime as a random variable $T_k \sim Exp(\nu)$ with the post-release life expectancy $\frac{1}{\nu}$. 
		\item We calculate the distances between $N_0$ individuals and the centers of $n_P$ traps at each time step, then store them in a matrix $dist$ of dimension $n_P \times n \times N_0$ where 
		
		\begin{center}
			$dist(i,s,k) = $ distance between $X_{s\Delta t}^k$ and $q_i$.
		\end{center}
		\item For the individual $k$, we take the {\bf first} step $s$ such that it belongs to some trap $\omega_i$ (that is, $dist(i, s, k) \leq R$). When a trajectory $X_t$ belongs to some $\omega_i$, the probability that the mosquito is captured between two timesteps $t$ and $t+dt$, conditionally on the fact that it has not been captured before is
		\begin{equation}
			\gamma \int_t^{t+dt} e^{-f_i(X_t) \, \tau} \, d\tau/e^{-f_i(X_t)\,t}=1-e^{-f_i(X_t) \, \Delta t}\approx f_i(X_t) \, \Delta t.
		\end{equation}
		
		Take a number $c$ uniformly distributed in $[0,1]$, if $c$ is smaller than the above probability and the time $s\Delta t$ is smaller than the individual's lifetime $T_k$, it is captured. If it is not captured in this step, we find the next step that satisfies this condition. 
		\item We count the number of mosquitoes captured in trap $\omega_i$ on day $j$ and obtain the simulated observation $\{\Tilde{y}_i^j \}$. 
	\end{itemize}

\end{document}